\documentclass{webofc}
\usepackage[varg]{txfonts}   
\usepackage{combelow}
\usepackage{multicol}
\usepackage{wrapfig,booktabs}
\usepackage{titlesec}
\renewcommand{\d}{\mathrm{d}}
\newcommand{\epn}{\frac{\d E_\perp^{(0)}}{\d \eta}}
\begin{document}
\title{Establishing the Range of Applicability of Hydrodynamics in High-Energy Collisions }
%
%

\author{\firstname{Clemens} \lastname{Werthmann}\inst{1,3}\fnsep\thanks{\email{cwerthmann@physik.uni-bielefeld.de}} \and
        \firstname{Victor E.} \lastname{Ambru\cb{s}}\inst{2}\fnsep \and
        \firstname{Sören} \lastname{Schlichting}\inst{1}\fnsep
}

\institute{Fakultät für Physik, Universität Bielefeld, D-33615 Bielefeld, Germany
\and
           Department of Physics, West University of Timi\cb{s}oara, Bd.~Vasile P\^arvan 4, Timi\cb{s}oara 300223, Romania
\and
           Incubator of Scientific Excellence-Centre for Simulations of Superdense Fluids, University of Wrocław, pl. Maxa Borna 9, 50-204 Wrocław, Poland
          }

\abstract{%
  We simulate the space-time dynamics of high-energy collisions based on a microscopic kinetic description, in order to determine the range of applicability of an effective description in relativistic viscous hydrodynamics. We find that hydrodynamics provides a quantitatively accurate description of collective flow when the average inverse Reynolds number $\mathrm{Re}^{-1}$ is sufficiently small and the early pre-equilibrium stage is properly accounted for. By determining the breakdown of hydrodynamics as a function of system size and energy, we find that it is quantitatively accurate in central lead-lead collisions at LHC energies, but should not be used in typical proton-lead or proton-proton collisions, where the development of collective flow cannot accurately be described within hydrodynamics.
}
\maketitle
\section{Introduction}\label{intro} The theoretical description of heavy ion collisions usually relies on hydrodynamics as one of its main components. However, the applicability of hydrodynamics on nucleonic time- and length-scales is unclear, as it requires a scale separation between microscopic degrees of freedom and the system size as well as some degree of equilibration. The discovery of collective flow in small systems~\cite{Nagle:2018nvi} has led to attempts to describe even these in hydrodynamic simulations, which calls for a critical examination of its validity.

Kinetic theory is a microscopic description that is applicable to dilute and far-from-equilibrium systems and converges to hydrodynamics in the limit of high interaction rate close to equilibrium. We investigate the accuracy of hydrodynamic results as a function of evolution time and system size by comparing to kinetic theory on the basis of observables related to transverse flow~\cite{Ambrus:2022koq,Ambrus:2022qya}.

\section{Setup}

We consider 2+1D simulations of the time evolution of an initial distribution with vanishing transverse anisotropy. The initial energy density profile was obtained as an average of events from the 30-40\,\% centrality class of Pb+Pb collisions at 5.02\,TeV (see \cite{Borghini:2022iym} for details). 

In kinetic theory, we describe the system as a single phase space distribution $f$ of massless on-shell bosons. The time evolution is given by the Boltzmann equation in conformal relaxation time approximation (RTA).
\begin{equation}
        p^\mu \partial_\mu f =C_{\rm RTA}[f]=- \frac{p^\mu u_\mu}{\tau_R} (f-f_{\rm eq}) \ , \ \ \  \tau_R=5\frac{\eta}{s}T^{-1}
\end{equation}
While this simplified setup will not provide a quantitatively accurate description of experimental data, it does allow for a detailed study of the qualitative behaviour of hadronic collision systems. One of the main advantages is that the dynamics will depend only on the geometry of the initial state and a single dimensionless parameter, the opacity $\hat{\gamma}$ \cite{Kurkela:2019kip}. It quantifies the total interaction rate in the system and encodes dependences on the shear viscosity ${\eta}/{s}$, the initial transverse radius $R$ and initial energy scale $\d E_\perp^{0}/\d \eta$. 
\begin{equation}\hat{\gamma} =\left(5{\frac{\eta}{s}}\right)^{-1}\left(\frac{1}{a\pi} {R}{\epn} \right)^{1/4}
\label{eq:ghat}
\end{equation}
In practice, we vary the shear viscosity ${\eta}/{s}$, but this scaling argument ensures that this is equivalent to varying system size.

The setup of hydrodynamics has to be chosen carefully such as to enable a reasonable comparison to kinetic theory. Consequently, we employ a conformal equation of state and choose the hydrodynamic transport coefficients to reproduce the behaviour of conformal RTA. Furthermore, hydrodynamics can not be expected to agree with kinetic theory in the early far-from-equilibrium stage. We therefore initialize the hydrodynamic simulations in a hybrid setup using profiles of the energy-momentum tensor from a kinetic theory simulation after it has partially equilibrated. The degree of equilibration can be assessed in terms of the inverse Reynolds number $\mathrm{Re}^{-1}=\left({6\pi_{\mu\nu}\pi^{\mu\nu}}/{e^2}\right)^{1/2}$\;\cite{Denicol:2012cn}, where $e$ is the Landau restframe energy density and $\pi^{\mu\nu}$ is the shear stress tensor. Thus, we decide when to start hydrodynamics based on the average value of this quantity weighted with the energy density, which we denote as $\left\langle\mathrm{Re}^{-1}\right\rangle_\epsilon$. Our simulations were carried out in vHLLE \cite{Karpenko:2013wva}. The specific setup is detailed in Sec. IIIC of~\cite{Ambrus:2022koq}.

    
        
            
    

\begin{figure}[b]
    \centering
    \includegraphics[width=0.313\textwidth]{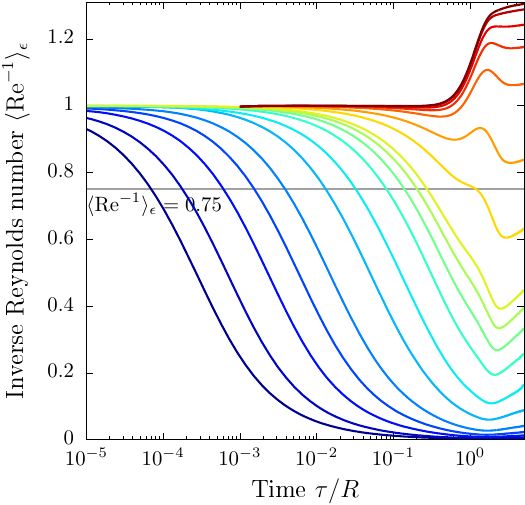}
    \includegraphics[width=0.325\textwidth]{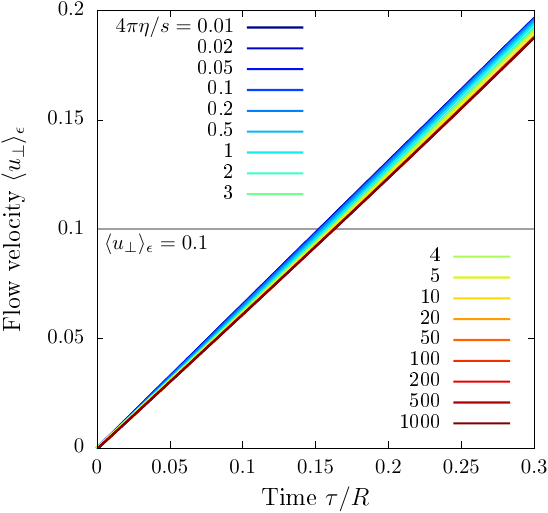}
    \includegraphics[width=0.32\textwidth]{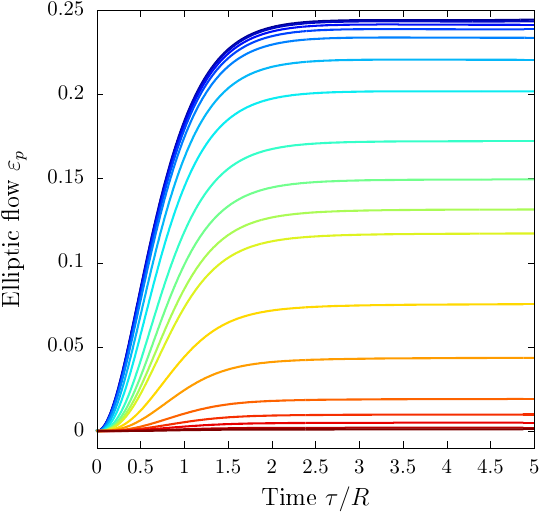}
    \caption{Time evolution of (left) inverse Reynolds number, (middle) transverse flow velocity and (right) elliptic flow for opacities ranging across four orders of magnitude. See Eq.~(\ref{eq:ghat}) for the relation between opacity and shear viscosity.}
    \label{fig:tevo}
    \vspace{-20pt}
\end{figure}

\section{Equilibration and onset of transverse dynamics}

To understand where hydrodynamics fails to accurately describe the system, we first study the time evolution of some observables of interest. Fig. \ref{fig:tevo} shows the time evolution in kinetic theory for opacities ranging across four orders of magnitude. The left plot shows the time evolution of $\left\langle\mathrm{Re}^{-1}\right\rangle_\epsilon$, which measures the degree of departure from equilibrium. Its behaviour is strongly dependent on the opacity, indicating that large systems equilibrate on a timescale that is a small fraction of their size, while some small systems never fully equilibrate.

The time evolution of the mean transverse flow velocity $\left\langle u_\perp\right\rangle_\epsilon$ is linear at early times. As this is mostly driven by the transverse geometry, its timescale has negligible dependence on the opacity. All systems reach a mean transverse flow velocity of $\left\langle u_\perp\right\rangle_\epsilon=0.1$ at roughly $\tau\approx 0.15 R$, which we use as a criterion for the onset of transverse expansion.

The buildup of elliptic flow $\varepsilon_p$ is driven by transverse expansion. The right plot shows that it is mostly built up around $\tau \sim R$. Qualitatively the time dependence curves are similar for all opacities, but their magnitude varies from no flow in the free-streaming limit to some large opacity limiting value $\epsilon_p\approx 0.25$, which might indicate that these systems follow the behaviour of ideal hydrodynamics. If one wanted to describe only elliptic flow at large opacities, one might come to the conclusion that the pre-equilibrium behaviour is irrelevant and hydrodynamics is always accurate independent of its initialization time. However, this is not true, as the transverse geometry - including eccentricities - will be modified during the pre-equilibrium cooling~\cite{Ambrus:2021fej}.


\begin{figure}[b]
\centering
    \includegraphics[width=.46\textwidth]{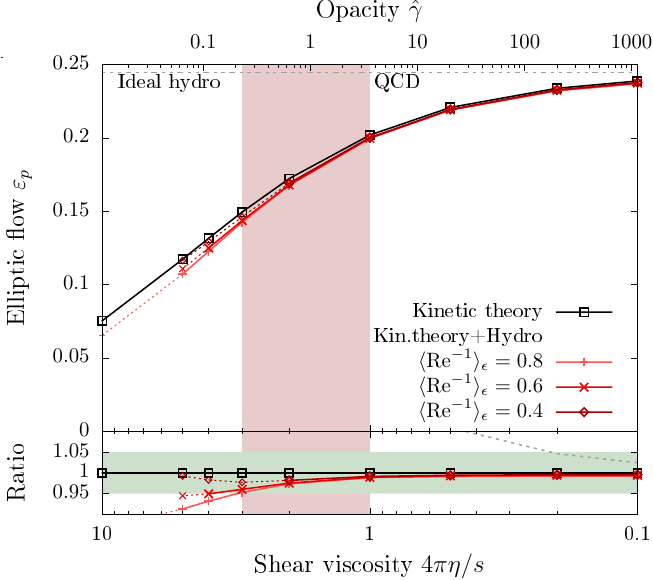}
    \includegraphics[width=.53\textwidth]{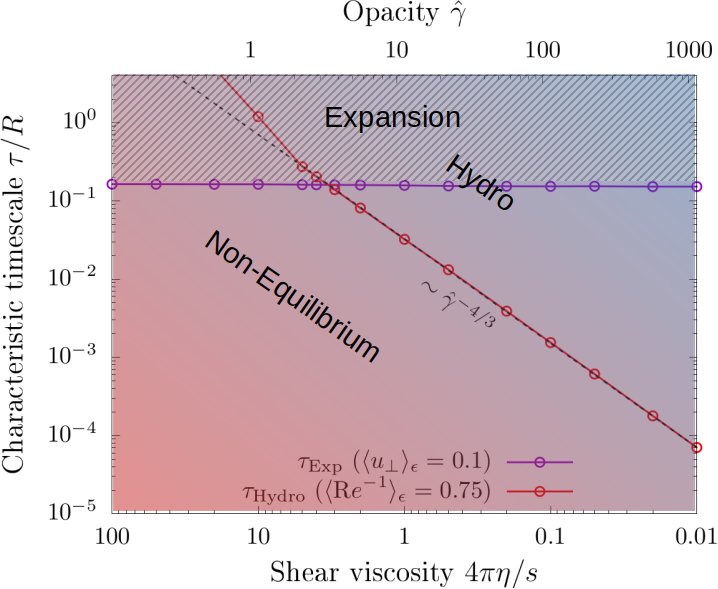}
    \caption{Left: Comparison of final values ($\tau=3R$) of elliptic flow in hybrid schemes at three different switching times (different shades of red) to kinetic theory (black) as a funciton of opacity. Right: Timescales of the onset of transverse expansion and hydrodynamization in kinetic theory simulations as a function of opacity. The purple line signifies the onset of transverse expansion, i.e. when a system is above the line it is subject to transverse expansion and flow develops. On the other hand, the pink line indicates a sufficient degree of equilibration for hyrodynamics to apply, i.e. to the left of this curve, the system is out-of-equilibrium whereas to the right of this line, it can be described accurately using hydrodynamics.}
    \label{fig:applicability}
\vspace{-20pt}
\end{figure}

\section{Applicability of Hydrodynamics}

We want to extract a criterion for the applicability of hydrodynamics based on the accuracy of late time results for transverse flow observables in hybrid schemes. The left plot in Fig. \ref{fig:applicability} shows a comparison of elliptic flow $\varepsilon_p$ results from kinetic theory and hybrid results employing three different switching criteria. The results improve continuously with decreasing value of $\langle \mathrm{Re}^{-1}\rangle_\epsilon$, i.e. with increasing degree of equilibration at the time of switching from kinetic theory to hydrodynamics. Evidently, this is because a smaller part of the pre-equilibrium period is simulated in hydrodynamics, which does not correctly describe the system's behaviour in this regime. Comparing also results for  $\frac{\d E_\perp}{d\eta}$ and $\langle u_T\rangle_\epsilon$ \cite{Ambrus:2022qya}, we find that all hybrid results show less than 5\% disagreement with kinetic theory if $\langle \mathrm{Re}^{-1}\rangle_\epsilon$ is below a critical value $\mathrm{Re}^{-1}_c\approx 0.75$, so we identify this as the criterion of applicability.

The right plot of Fig. \ref{fig:applicability} now compares our extracted characteristic timescales of the onset of transverse expansion and hydrodynamization as a function of opacity. These can also be taken from the left and middle plots of Fig. \ref{fig:tevo}. While the former has little opacity dependence, hydrodynamization takes much longer for smaller systems, if it even sets in at all. This means that for small opacities, the system has not hydrodynamized at the onset of transverse expansion, so it becomes necessary to employ a 2+1D nonequilibrium description in order to achieve accurate results. The smallest opacities for which hydrodynamics remains accurate are on the order of $\hat{\gamma}\approx 3$, corresponding to the crossing of the lines. This result could in principle be conditional to the specific geometry of the chosen initial condition. However, we also tried varying the centrality at fixed shear viscosity and found a similar criterion~\cite{Ambrus:2022qya}.

\begin{wraptable}{r}{6.6cm}
\centering
\begin{tabular}{c|c|c|c}
        System & $\frac{\d E^{0}_{\bot}}{\d\eta}$ & $R$  & $\hat{\gamma}$ \\ 
        &[GeV] & [fm] & \\
        \hline
         p+p (min. bias) & 7.1 & 0.12 & 0.70\\
         p+Pb (min.bias) & 24 & 0.81 & 1.5 \\
         p+Pb (high mult.) & 230 & 0.81 & 2.7 \\ 
         O+O (70-80) & 13 & 0.88 & 1.4 \\
         O+O (30-40) & 55 & 1.13 & 2.2 \\
         O+O (0-5) & 140 & 1.61 & 3.1 \\
         Pb+Pb (70-80) & 85.1 & 2.16 & 2.70\\
         Pb+Pb (30-40) & 1280 & 2.78 & 5.66 \\
         Pb+Pb (0-5) & 5670 & 3.94 & 8.97
    \end{tabular}
\caption{Opacity estimates for several collision systems}
\label{tab:opacities}       
\vspace{-10pt}
\end{wraptable}

In order to understand what this means for real collision systems that are examined at the LHC, we computed their typical opacity values 
using Eq.~(\ref{eq:ghat}), based on a value of the shear viscosity of $4\pi \eta / s = 2$ and estimates for the system energy and size.
The results are compiled in Table \ref{tab:opacities}. We find that Pb+Pb systems mostly behave hydrodynamically except in peripheral collisions. On the other hand, p+p collisions are far from hydrodynamic behaviour. p+Pb collisions come closer to the regime of applicability of hydrodynamics, but still do not reach it. But interestingly, O+O collisions might probe the transition regime to hydrodynamic behaviour.

\textbf{Acknowledgements:} This work is supported by the Deutsche Forschungsgemeinschaft (DFG, German Research Foundation) through the CRC-TR 211 ’Strong-interaction matter under extreme conditions’– project
number 315477589 – TRR 211. C.W. was supported by the program Excellence Initiative–Research University of the University of Wrocław of the Ministry of Education and Science.

%

\begin{thebibliography}{}
%
%
\bibitem{Nagle:2018nvi}
J.~L.~Nagle and W.~A.~Zajc,
Ann. Rev. Nucl. Part. Sci. \textbf{68} (2018) 211.
\bibitem{Ambrus:2022koq}
V.~E.~Ambrus, S.~Schlichting and C.~Werthmann,
Phys. Rev. D \textbf{107} (2023) 094013.
\bibitem{Ambrus:2022qya}
V.~E.~Ambrus, S.~Schlichting and C.~Werthmann,
Phys. Rev. Lett. \textbf{130} (2023) 152301.
\bibitem{Borghini:2022iym}
N.~Borghini, M.~Borrell, N.~Feld, H.~Roch, S.~Schlichting and C.~Werthmann,
Phys. Rev. C \textbf{107} (2023) 034905.
\bibitem{Kurkela:2019kip}
A.~Kurkela, U.~A.~Wiedemann and B.~Wu,
Eur. Phys. J. C \textbf{79} (2019) 965.

\bibitem{Denicol:2012cn}
    G.~S.~Denicol, H.~Niemi, E.~Molnar and D.~H.~Rischke,
    Phys. Rev. D \textbf{85} (2012) 114047.


\bibitem{Karpenko:2013wva}
I.~Karpenko, P.~Huovinen and M.~Bleicher,
Comput. Phys. Commun. \textbf{185} (2014) 3016.
\bibitem{Ambrus:2021fej}
V.~E.~Ambrus, S.~Schlichting and C.~Werthmann,
Phys. Rev. D \textbf{105} (2022) 014031.

\end{thebibliography}
%
%

\end{document}